\pdfoutput=1
\documentclass[submission, Proceedings]{SciPost}

\usepackage{lineno}
\usepackage{subcaption}
\usepackage{enumitem}
\usepackage{multirow}

\hypersetup{
    colorlinks,
    linkcolor={red!50!black},
    citecolor={blue!50!black},
    urlcolor={blue!80!black}
}

\usepackage[bitstream-charter]{mathdesign}
\DeclareSymbolFont{usualmathcal}{OMS}{cmsy}{m}{n}
\DeclareSymbolFontAlphabet{\mathcal}{usualmathcal}

\begin{document}

\begin{center}{\Large \textbf{
A new
Wilson Line-based classical action for gluodynamics\\
}}\end{center}

\begin{center}
Hiren Kakkad\textsuperscript{1},
Piotr Kotko\textsuperscript{1$\star$} and
Anna Stasto\textsuperscript{2}
\end{center}

\begin{center}
{\bf 1} {\it AGH University Of Science and Technology, Faculty of Physics and Applied Computer Science} \\ 
{\it Mickiewicza 30, 30-059 Krakow, Poland}
\\
{\bf 2} {\it The Pennsylvania State University, Physics Department}\\ 
{\it 104 Davey Lab, University Park, PA 16802, USA }
\\

* piotr.kotko@fis.agh.edu.pl
\end{center}

\begin{center}
October 15, 2021
\end{center}


\definecolor{palegray}{gray}{0.95}
\begin{center}
\colorbox{palegray}{
  \begin{tabular}{rr}
  \begin{minipage}{0.1\textwidth}
    \includegraphics[width=35mm]{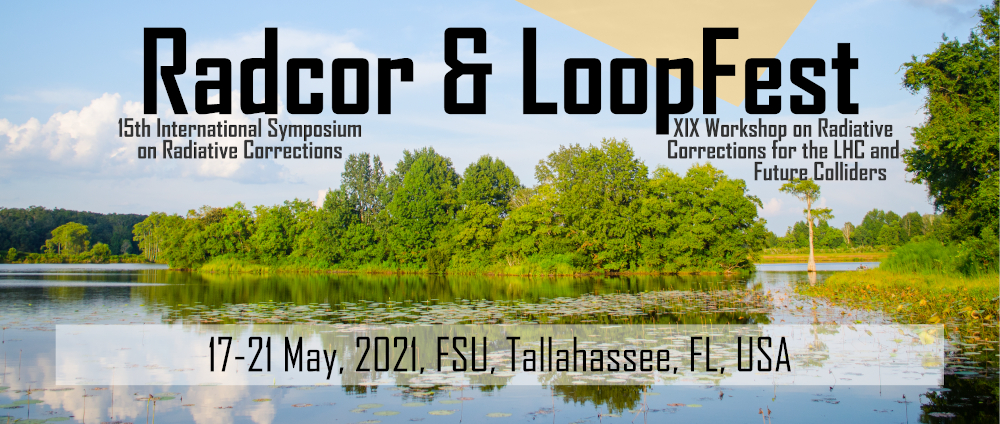}
  \end{minipage}
  &
  \begin{minipage}{0.85\textwidth}
    \begin{center}
    {\it 15th International Symposium on Radiative Corrections: \\Applications of Quantum Field Theory to Phenomenology,}\\
    {\it FSU, Tallahasse, FL, USA, 17-21 May 2021} \\
    \doi{10.21468/SciPostPhysProc.?}\\
    \end{center}
  \end{minipage}
\end{tabular}
}
\end{center}

\section*{Abstract}
{\bf

We develop a new classical action that in addition to $\mathrm{MHV}$ vertices contains also $\mathrm{N^kMHV}$ vertices, where $1\leq k \leq n-4$ with $n$ the number of external legs.
The lowest order vertex is the four-point MHV vertex -- there is no three point vertex and thus the amplitude calculation involves fewer vertices than in the CSW method and, obviously, considerably  fewer than in the standard Yang-Mills action.
The action is obtained by a canonical transformation of the Yang-Mills action in the light-cone gauge, where the field transformations are based on the Wilson line functionals.
}

\vspace{10pt}
\noindent\rule{\textwidth}{1pt}
\tableofcontents\thispagestyle{fancy}
\noindent\rule{\textwidth}{1pt}
\vspace{10pt}

\newpage

\section{Introduction}
\label{sec:intro}
 The following work focuses on a description of pure gluonic scattering amplitudes in terms of a new action, currently developed at the classical level (thus suitable for tree amplitudes). Despite considered as fundamental, gluon fields are often not the most efficient degrees of freedom for computing amplitudes. Interestingly, in \cite{Kotko2017}, the Maximally Helicity Violating (MHV) vertices used in the Cachazo-Svrcek-Witten (CSW) method \cite{Cachazo2004} were shown to be connected with straight infinite Wilson lines on certain complex plane (self-dual plane). These Wilson lines emerge as the transformation of the positive helicity field appearing in the light cone Yang-Mills action, to a new action (often called as the 'MHV action') where the MHV vertices are explicit \cite{Mansfield2006,Ettle2006b,Ettle2007,Ettle2008}. A similar Wilson line-type structure was found, in \cite{Kakkad2020}, for the negative helicity field. Moreover in the latter, we postulated, that it should be a part of a bigger structure, extending beyond the self-dual plane.

Indeed, in \cite{kakkad2021}, we derived a new classical action for gluodynamics in which the fields are directly related to Wilson line functionals extending over both the self-dual and the anti-self-dual planes. The action is most easily derived through a canonical transformation of the anti-self-dual part of the MHV action, but we also discuss a direct link between the new action and the Yang-Mills action. 
The key property of the new action is that it does not have the triple-gluon vertices at all. This is because the triple-gluon vertices have been effectively resummed inside the Wilson lines. Thus, the lowest multiplicity vertex is the four-point MHV vertex. Higher-point vertices include not only the MHV vertices, but also other helicity configurations. The number of diagrams needed to obtain amplitudes beyond the MHV level is thus greatly reduced. We performed explicit calculations within the new formulation of several higher multiplicity amplitudes, to verify the consistency of the results.

\section{MHV Lagrangian}

The starting point is the full Yang-Mills action on the constant light-cone time $x^+$ in the light-cone gauge $\hat{A}^+=0$. We denote $\hat{A}=A_at^a$, here $t^a$ are  color generators in the fundamental representation satisfying $\left[t^{a},t^{b}\right]=i\sqrt{2}f^{abc}t^{c}$ and $\mathrm{Tr}(t^{a} t^{b}) = \delta^{ab}$. Integrating out the $\hat{A}^-$ fields (appearing quadratically) from the partition function, leaves only two  complex fields $\hat{A}^{\bullet}$, $\hat{A}^{\star}$ that correspond to plus-helicity and minus-helicity gluon fields. We use the so-called 'double-null' coordinates defined as 
$v^{+}=v\cdot\eta$, $v^{-}=v\cdot\tilde{\eta}$, 
$v^{\bullet}=v\cdot\varepsilon_{\bot}^{+}$,  
$v^{\star}=v\cdot\varepsilon_{\bot}^{-}$
with the two light-like basis four-vectors 
$\eta=\left(1,0,0,-1\right)/\sqrt{2}$, $\tilde{\eta}=\left(1,0,0,1\right)/\sqrt{2}$,
and two space like complex four-vectors spanning the transverse plane
$\varepsilon_{\perp}^{\pm}=\frac{1}{\sqrt{2}}\left(0,1,\pm i,0\right)$. The Yang-Mills action in this setup reads
\begin{multline}
S_{\mathrm{Y-M}}^{\left(\mathrm{LC}\right)}\left[A^{\bullet},A^{\star}\right]=\int dx^{+}\int d^{3}\mathbf{x}\,\,\Bigg\{ 
-\mathrm{Tr}\,\hat{A}^{\bullet}\square\hat{A}^{\star}
-2ig\,\mathrm{Tr}\,\partial_{-}^{-1}\partial_{\bullet} \hat{A}^{\bullet}\left[\partial_{-}\hat{A}^{\star},\hat{A}^{\bullet}\right] \\
-2ig\,\mathrm{Tr}\,\partial_{-}^{-1}\partial_{\star}\hat{A}^{\star}\left[\partial_{-}\hat{A}^{\bullet},\hat{A}^{\star}\right]
-2g^{2}\,\mathrm{Tr}\,\left[\partial_{-}\hat{A}^{\bullet},\hat{A}^{\star}\right]\partial_{-}^{-2}\left[\partial_{-}\hat{A}^{\star},\hat{A}^{\bullet}\right]
\Bigg\}
\,,\label{eq:YM_LC_action}
\end{multline}
where $\square=2(\partial_+\partial_- - \partial_{\bullet}\partial_{\star})$. Thus, we see there are $(++-)$, $(--+)$ and $(++-\,-)$ vertices in the action. Above, the bold position vector is defined as $\mathbf{x}\equiv\left(x^{-},x^{\bullet},x^{\star}\right)$. 

The MHV action \cite{Mansfield2006}, implementing the CSW rules \cite{Cachazo2004} is obtained from the Yang-Mills action Eq. (\ref{eq:YM_LC_action}) by canonically transforming both the fields to a new pair of fields ($\hat{B}^{\bullet}$, $\hat{B}^{\star}$) with a requirement that the kinetic term and the $(++-)$ triple-gluon vertex in Eq. (\ref{eq:YM_LC_action}) is mapped to the kinetic term in the new action:
\begin{equation}
\mathrm{Tr}\,\hat{A}^{\bullet}\square\hat{A}^{\star}
+2ig\,\mathrm{Tr}\,\partial_{-}^{-1}\partial_{\bullet} \hat{A}^{\bullet}\left[\partial_{-}\hat{A}^{\star},\hat{A}^{\bullet}\right]
\,\, \longrightarrow \,\,
\mathrm{Tr}\,\hat{B}^{\bullet}\square\hat{B}^{\star}
\,.\label{eq:MansfieldTransf1}
\end{equation}
Solving the above transformation for $\hat{A}^{\bullet}$, $\hat{A}^{\star}$ and substituting it in Eq. (\ref{eq:YM_LC_action}) results in the MHV action consisting of an infinite set of MHV vertices
\begin{equation}
    S_{\mathrm{Y-M}}^{\left(\mathrm{LC}\right)}\left[{B}^{\bullet}, {B}^{\star}\right]=\int dx^{+}\left(
-\int d^{3}\mathbf{x}\,\mathrm{Tr}\,\hat{B}^{\bullet}\square\hat{B}^{\star} 
+\mathcal{L}_{-\,-\,+}^{\left(\mathrm{LC}\right)}+\dots +\mathcal{L}_{-\,-\,+\,\dots\,+}^{\left(\mathrm{LC}\right)}+\dots\right)\,,\label{eq:MHV_action}
\end{equation}
where $\mathcal{L}_{-\,-\,+\,\dots\,+}^{\left(\mathrm{LC}\right)}$ represents a generic $n$-point MHV vertex in the action, which in our conventions has the following form in the momentum space
\begin{multline}
\mathcal{L}_{-\,-\,+\,\dots\,+}^{\left(\mathrm{LC}\right)}=\int d^{3}\mathbf{p}_{1}\dots d^{3}\mathbf{p}_{n}\delta^{3}\left(\mathbf{p}_{1}+\dots+\mathbf{p}_{n}\right)\,
\widetilde{\mathcal{V}}_{-\,-\,+\,\dots\,+}^{b_{1}\dots b_{n}}\left(\mathbf{p}_{1},\dots,\mathbf{p}_{n}\right)\\
\widetilde{B}_{b_{1}}^{\star}\left(x^+;\mathbf{p}_{1}\right)\widetilde{B}_{b_{2}}^{\star}\left(x^+;\mathbf{p}_{2}\right)\widetilde{B}_{b_{3}}^{\bullet}\left(x^+;\mathbf{p}_{3}\right)\dots\widetilde{B}_{b_{n}}^{\bullet}\left(x^+;\mathbf{p}_{n}\right)
\,,
\label{eq:MHV_n_point}
\end{multline}
with the MHV vertices 
\begin{equation}
\widetilde{\mathcal{V}}_{-\,-\,+\,\dots\,+}^{b_{1}\dots b_{n}}\left(\mathbf{p}_{1},\dots,\mathbf{p}_{n}\right)=  \sum  \mathrm{Tr}\left(t^{b_1}\dots t^{b_n}\right)
\frac{(-g)^{n-2}}{(n-2)!}  \left(\frac{p_{1}^{+}}{p_{2}^{+}}\right)^{2}
\frac{\widetilde{v}_{21}^{*4}}{\widetilde{v}_{1n}^{*}\widetilde{v}_{n\left(n-1\right)}^{*}\widetilde{v}_{\left(n-1\right)\left(n-2\right)}^{*}\dots\widetilde{v}_{21}^{*}}
\,,
\label{eq:MHV_vertex}
\end{equation}
where the sum is over noncyclic permutations. Above, we introduced spinor-like variables
\begin{equation}
    \tilde{v}_{ij}=
    p_i^+\left(\frac{p_{j}^{\star}}{p_{j}^{+}}-\frac{p_{i}^{\star}}{p_{i}^{+}}\right), \qquad 
\tilde{v}^*_{ij}=
    p_i^+\left(\frac{p_{j}^{\bullet}}{p_{j}^{+}}-\frac{p_{i}^{\bullet}}{p_{i}^{+}}\right)\, .
\label{eq:vtilde}
\end{equation}
The $\tilde{v}_{ij}$, $\tilde{v}_{ij}^*$ symbols are directly proportional to the spinor products $\left< ij \right>$ and $\left[ ij \right]$.

\section{Wilson lines in MHV Lagrangian}
\begin{figure}
\vspace{-0.3cm}
\centering
\vspace{-0.3cm}
\begin{subfigure}{.5\textwidth}
  \centering
  \includegraphics[width=.7\linewidth]{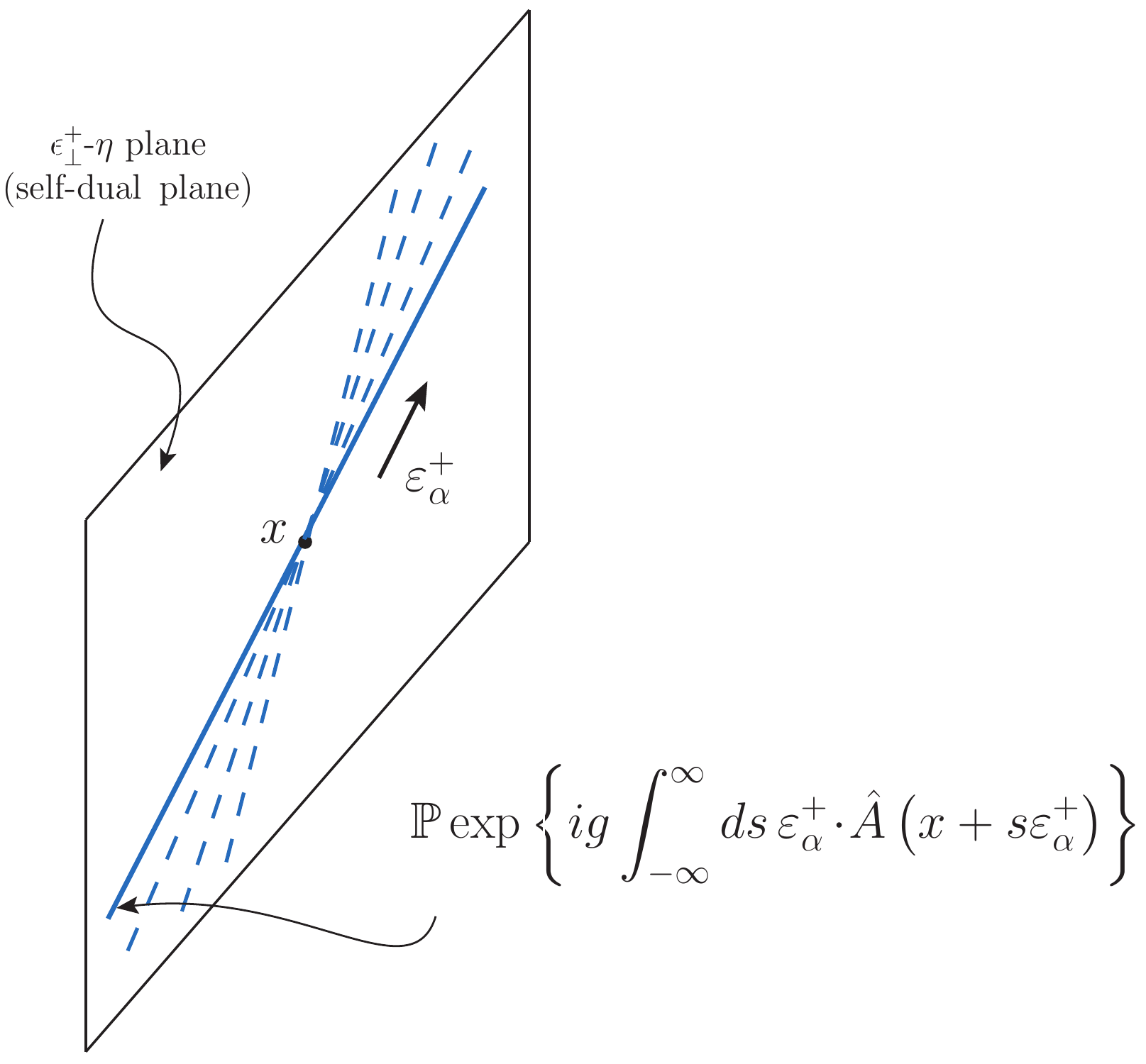}
  \caption{}
  \label{fig:SDplane}
\end{subfigure}%
\begin{subfigure}{.5\textwidth}
  \centering
  \includegraphics[width=.35\linewidth]{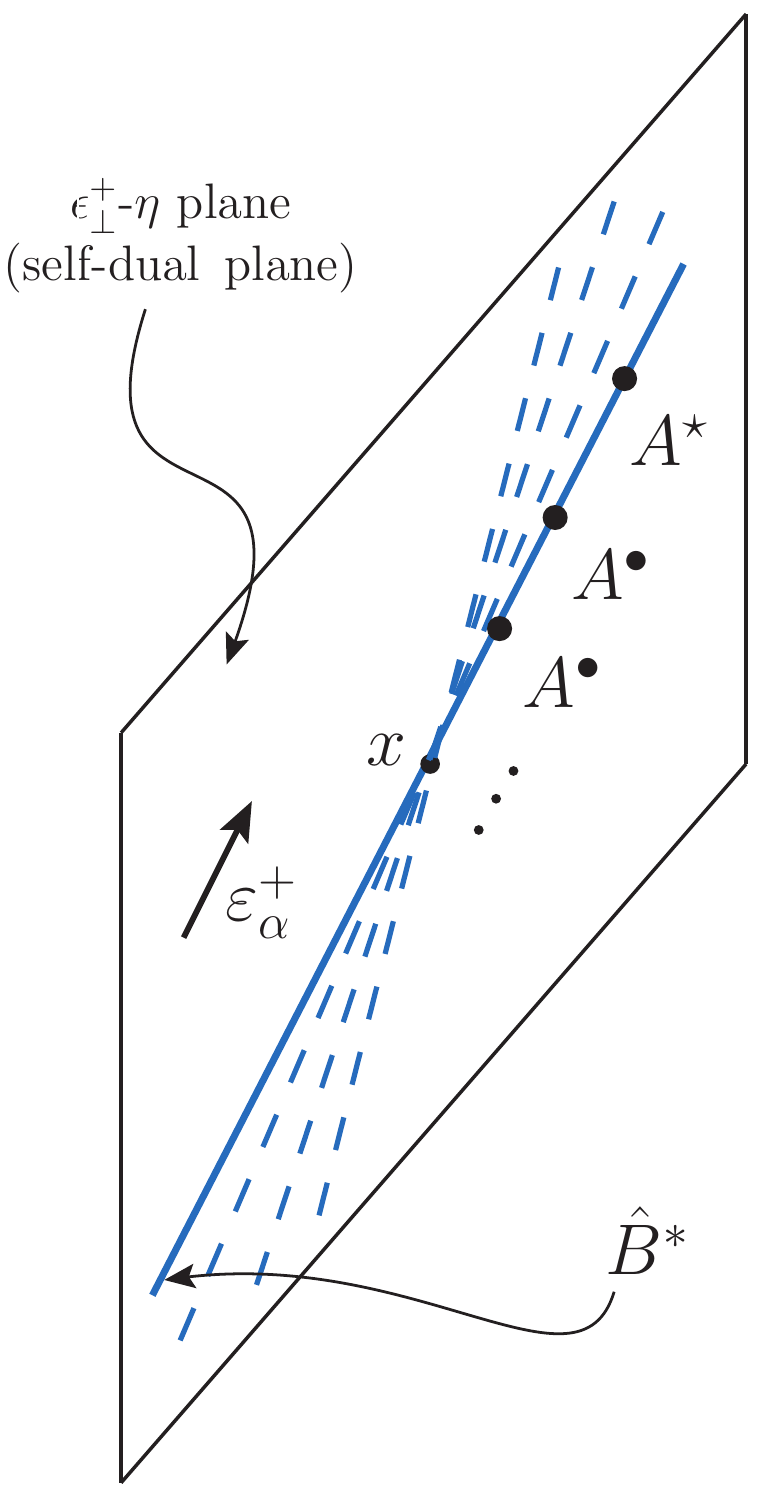}
  \caption{}
  \label{fig:SDplane_Bstar}
\end{subfigure}
\vspace{-0.3cm}
\caption{\scriptsize Left:
    The straight infinite Wilson line $B^{\bullet}$ lying on the plane spanned by $\varepsilon^+_{\alpha}=\varepsilon^+_{\perp}-\alpha\eta$  
    and integrated over all $\alpha$ (the dashed lines represent tilted Wilson lines due to the change of $\alpha$).  Right: The $B^{\star}$ field can be represented as the straight infinite Wilson line similar to the one on the left, but where one $A^{\bullet}$ field has been replaced by the $A^{\star}$ field in the functional expansion (with a suitable symmetry factor).}
\label{fig:B_fields}
\end{figure}

In the original work \cite{Mansfield2006}, the action was constructed using only analytic properties of the transformations and equivalence
theorem for the S-matrix. The explicit solution for $\hat{A}^{\bullet}$ and $\hat{A}^{\star}$ fields, in momentum
space, was found in \cite{Ettle2006b}.  The Wilson line interpretation of the new fields in the MHV action was first discussed in \cite{Kotko2017} where the plus helicity field, $B^{\bullet}_a[\hat{A}^{\bullet}](x)$, was shown to be the straight infinite Wilson line $B^{\bullet}_a[A^{\bullet}](x)=\mathcal{W}_{(+)}^a[A](x)$, where  for a generic vector field $K^{\mu}$ we defined
\begin{equation}
     \mathcal{W}^{a}_{(\pm)}[K](x)=\int_{-\infty}^{\infty}d\alpha\,\mathrm{Tr}\left\{ \frac{1}{2\pi g}t^{a}\partial_{-}\, \mathbb{P}\exp\left[ig\int_{-\infty}^{\infty}\! ds\, \varepsilon_{\alpha}^{\pm}\cdot \hat{K}\left(x+s\varepsilon_{\alpha}^{\pm}\right)\right]\right\} \, ,
\label{eq:WL_gen}
\end{equation}
with $\varepsilon_{\alpha}^{\pm\, \mu} = \varepsilon_{\perp}^{\pm\, \mu }- \alpha \eta^{\mu} \, $. Notice, that the latter four vector has the form of a gluon polarization vector. Indeed for $\alpha=p\cdot\varepsilon_{\perp}^{\pm}/p^{+}$, it is the transverse polarization vector for a gluon with momentum $p$. Thus, in momentum space the Wilson line $B^{\bullet}_a[\hat{A}^{\bullet}](x)$ lies along the plus helicity polarization vector.
Interestingly, the two vectors defining the direction of the Wilson line, $\varepsilon_{\perp}^+$ and $\eta$, span the so-called self-dual plane (the plane on which the tensors are self-dual). Note, however, that the Wilson line is integrated over all possible directions $\alpha$ on the self-dual plane gaining a projective character (see Fig.~\ref{fig:SDplane}).

The minus helicity field $ B_a^{\star}[\hat{A}^{\bullet},\hat{A}^{\star}](x)$, on the other hand, was shown in \cite{Kakkad2020} to be a similar Wilson line, but with an insertion of the minus helicty gluon field at certain point on the line (see Fig.~\ref{fig:SDplane_Bstar}), more precisely
\begin{equation}
    B_a^{\star}[A^{\bullet},A^{\star}](x) = 
    \int\! d^3\mathbf{y} \,
     \left[ \frac{\partial^2_-(y)}{\partial^2_-(x)} \,
     \frac{\delta \mathcal{W}^a_{(+)}[A](x^+;\mathbf{x})}{\delta {A}_c^{\bullet}(x^+;\mathbf{y})} \right] 
     {A}_c^{\star}(x^+;\mathbf{y})
      \, ,
    \label{eq:Bstar_Bbullet}
\end{equation}
where $\partial_-(x)=\partial/\partial x^-$. 
Since it is natural to think about the $A^{\star}$ fields as belonging to Wilson lines living within the anti-self-dual plane spanned by $\varepsilon^-_{\alpha}$ and $\eta$
(recall that the $B^{\bullet}$ lives on the plane spanned by $\varepsilon^+_{\alpha}$), we conjectured in \cite{Kakkad2020} that the solution (\ref{eq:Bstar_Bbullet}) should just be a cut through a bigger structure, spanning over both planes.

\section{New classical action}
\label{sec:Z_action}

The canonical transformation, Eq.  \eqref{eq:MansfieldTransf1}, eliminates one of the triple gloun vertex $(+ + -)$  while the other triple gloun vertex $(+ - -)$ still exists in the MHV action. Triple point vertices are not very effective building blocks for calculating amplitudes, and, actually they do not constitute any physical amplitude themselves -- in the on-shell limit they are zero (for real momenta). 
Motivated by the geometric considerations mentioned before and the above arguments we proposed in \cite{kakkad2021} another set of field transformations that lead to a new action. 

\subsection{Field Transformation}
\begin{figure}
\vspace{-0.3cm}
    \centering
    \includegraphics[width=8cm]{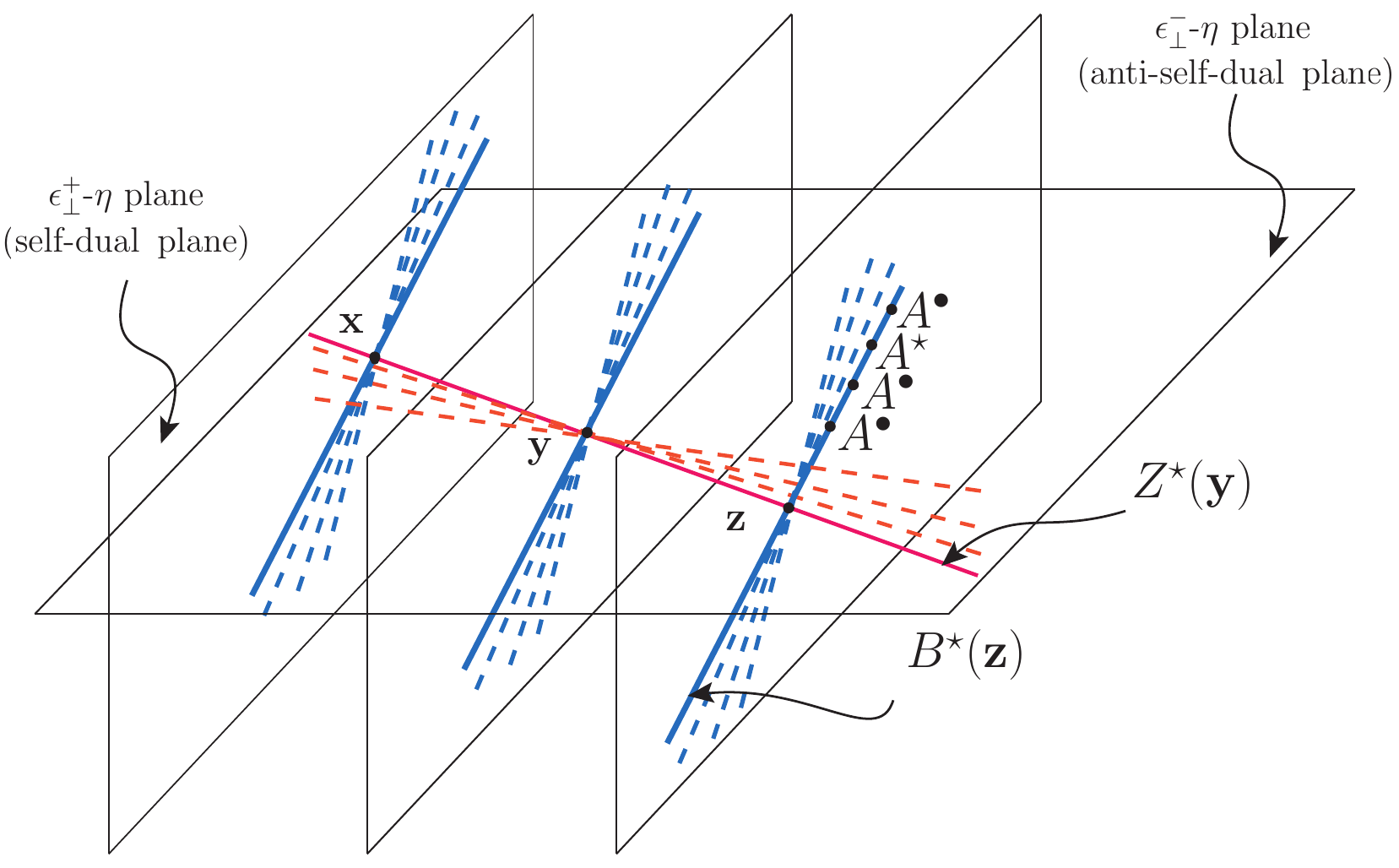}
    \caption{\scriptsize
    Schematic presentation of the geometric structure of the $Z^{\star}$ field (the structure of $Z^{\bullet}$ is quite similar). $Z^{\star}$ field is a Wilson line (with exactly the same analytic form as $B^{\bullet}$) of only $B^{\star}$ fields on anti-self-dual plane. Notice, each vertical plane is self-dual plane with $B^{\star}$ embedded in it as was showin in Fig.~\ref{fig:SDplane_Bstar}. 
    }
    \label{fig:geometry}
\end{figure}

The new canonical field transformations are based on path ordered exponentials of the gauge fields, extending over both the self-dual and anti-self-dual planes \cite{kakkad2021}:
\begin{equation}
    \left\{\hat{A}^{\bullet},\hat{A}^{\star}\right\} \rightarrow \Big\{\hat{Z}^{\bullet}\big[{A}^{\bullet},{A}^{\star}\big],\hat{Z}^{\star}\big[{A}^{\bullet},{A}^{\star}\big]\Big\} \, ,
    \label{eq:general_transf}
\end{equation}
It maps the kinetic term and both the triple-gluon vertices of the Yang-Mills action to a free term in the new action. In order to preserve the functional measure in the partition function, up to a field independent factor, it is necessary that the transformation is canonical. Although the transformation \eqref{eq:general_transf} is rather complicated, we found that, quite amazingly, the generating functional  $\mathcal{G}[A^{\bullet},Z^{\star}]$ for the transformation can be written in the following simple form:
\begin{equation}
    \mathcal{G}[A^\bullet,Z^\star](x^+) = - \int d^3\mathbf{x}\,\,\,\mathrm{Tr}\,
     \hat{\mathcal{W}}^{\,-1}_{(-)}[Z](x)\,\,
     \partial_- \hat{\mathcal{W}}_{(+)}[A](x) \,,
    \label{eq:generatingfunc3}
\end{equation}
The Yang-Mills and the new fields are related as:
\begin{equation}
     \partial_{-}A^{\star}_a(x^+,\mathbf{y}) =  \frac{\delta \, \mathcal{G}[A^{\bullet},Z^{\star} ](x^+)}{\delta A_a^{\bullet}\left(x^+,\mathbf{y}\right)} \,, \qquad 
     \partial_{-}Z^{\bullet}_a(x^+,\mathbf{y}) = - \frac{\delta \, \mathcal{G}[A^{\bullet},Z^{\star} ](x^+)}{\delta Z_a^{\star}\left(x^+,\mathbf{y}\right)} \,,
     \label{eq:generatingfunc2}
\end{equation}
In \cite{kakkad2021} we showed that the transformation  \eqref{eq:generatingfunc3} is equivalent to two canonical transformations: first transforming the self-dual part of the Yang-Mills action to the kinetic term in MHV action (\ref{eq:MansfieldTransf1}), and then transforming the anti-self-dual part in the latter to kinetic term in the new action
\begin{equation}
\mathcal{L}_{-\,+}[B^{\bullet},B^{\star}]+\mathcal{L}_{-\,-\,+}[B^{\bullet},B^{\star}]
\,\, \longrightarrow \,\,
\mathcal{L}_{-\,+}[Z^{\bullet},Z^{\star}]
\,,
\label{eq:BtoZtransform}
\end{equation}
Following this, the solution for $Z$ fields reads (see Fig.~\ref{fig:geometry})
\begin{align}
      & Z^{\star}_a[B^{\star}](x) =\mathcal{W}_{(-)}^a[B](x)\,, \nonumber \\   & Z_a^{\bullet}[B^{\bullet},B^{\star}](x) =
    \int\! d^3\mathbf{y} \,
     \left[ \frac{\partial^2_-(y)}{\partial^2_-(x)} \,
     \frac{\delta \mathcal{W}^a_{(-)}[B](x^+;\mathbf{x})}{\delta {B}_c^{\star}(x^+;\mathbf{y})} \right] 
     {B}_c^{\bullet}(x^+;\mathbf{y})
      \, .
      \label{eq:Zfield_transform}
\end{align}

\subsection{Structure of the action}

The new action can be most easily derived \cite{kakkad2021} by substituting the inverse of $Z$ fields (\ref{eq:Zfield_transform}) in the MHV action. For a generic Wilson line $\mathcal{W}$ (\ref{eq:WL_gen}), the inverse functional, $\mathcal{W}^{-1}$, is defined 
using the relation
$\mathcal{W}[\mathcal{W}^{-1}[K]]=K$. Using this, for the $B^{\star}$ field we find
\begin{equation}
    \widetilde{B}^{\star}_a(x^+;\mathbf{P}) = \sum_{n=1}^{\infty} 
    \int d^3\mathbf{p}_1\dots d^3\mathbf{p}_n \, \overline{\widetilde{\Psi}}\,^{a\{b_1\dots b_n\}}_n(\mathbf{P};\{\mathbf{p}_1,\dots ,\mathbf{p}_n\}) \prod_{i=1}^n\widetilde{Z}^{\star}_{b_i}(x^+;\mathbf{p}_i)\,,
    \label{eq:BstarZ_exp}
\end{equation}
with
\begin{equation}
    \overline{\widetilde \Psi}\,^{a \left \{b_1 \cdots b_n \right \}}_{n}(\mathbf{P}; \left \{\mathbf{p}_{1},  \dots ,\mathbf{p}_{n} \right \}) =- (-g)^{n-1} \,\,  
    \frac{{\widetilde v}_{(1 \cdots n)1}}{{\widetilde v}_{1(1 \cdots n)}} \, 
    \frac{\delta^{3} (\mathbf{p}_{1} + \cdots +\mathbf{p}_{n} - \mathbf{P})\,\,  \mathrm{Tr} (t^{a} t^{b_{1}} \cdots t^{b_{n}})}{{\widetilde v}_{21}{\widetilde v}_{32} \cdots {\widetilde v}_{n(n-1)}}  
      \, .
    \label{eq:psiBar_kernel}
\end{equation}
The expansion for the $B^{\bullet}$ field reads
\begin{equation}
     \widetilde{B}^{\bullet}_a(x^+;\mathbf{P}) = \sum_{n=1}^{\infty} 
    \int d^3\mathbf{p}_1\dots d^3\mathbf{p}_n \, \overline{\widetilde \Omega}\,^{a b_1 \left \{b_2 \cdots b_n \right \}}_{n}(\mathbf{P}; \mathbf{p_1} ,\left \{ \mathbf{p_2} , \dots ,\mathbf{p_n} \right \}) \widetilde{Z}^{\bullet}_{b_1}(x^+;\mathbf{p}_1)\prod_{i=2}^n\widetilde{Z}^{\star}_{b_i}(x^+;\mathbf{p}_i)\,,
    \label{eq:BbulletZ_exp}
\end{equation}
where
\begin{equation}
    \overline{\widetilde \Omega}\,^{a b_1 \left \{b_2 \cdots b_n \right \}}_{n}(\mathbf{P}; \mathbf{p}_{1} , \left \{ \mathbf{p}_{2} , \dots ,\mathbf{p}_{n} \right \} ) = n \left(\frac{p_1^+}{p_{1\cdots n}^+}\right)^2 \overline{\widetilde \Psi}\,^{a b_1 \cdots b_n }_{n}(\mathbf{P}; \mathbf{p}_{1},  \dots ,\mathbf{p}_{n}) \, .
    \label{eq:omegaBar_kernel}
\end{equation}
Upon substitution of the expansions \eqref{eq:BstarZ_exp}-\eqref{eq:BbulletZ_exp}, in the MHV action, we obtain the following generic structure of the new action: 
\begin{align}
S_{\mathrm{Y-M}}^{\left(\mathrm{LC}\right)}\left[Z^{\bullet},Z^{\star}\right] =\int dx^{+} \Bigg\{ & 
-\int d^{3}\mathbf{x}\,\mathrm{Tr}\,\hat{Z}^{\bullet}\square\hat{Z}^{\star} \nonumber \\
 & + \mathcal{L}^{(\mathrm{LC})}_{-\,-\,+\,+}+ \mathcal{L}^{(\mathrm{LC})}_{-\,-\,+\,+\,+}+\mathcal{L}^{(\mathrm{LC})}_{-\,-\,+\,+\,+\,+} + \dots \nonumber \\
& + \mathcal{L}^{(\mathrm{LC})}_{-\,-\,-\,+\,+}+ \mathcal{L}^{(\mathrm{LC})}_{-\,-\,-\,+\,+\,+}+\mathcal{L}^{(\mathrm{LC})}_{-\,-\,-\,+\,+\,+\,+} + \dots \nonumber \\
& \,\, \vdots \nonumber \\
& + \mathcal{L}^{(\mathrm{LC})}_{-\,-\,-\,\dots \,-\,+\,+}+ \mathcal{L}^{(\mathrm{LC})}_{\,-\,-\,-\,\dots\, -\,+\,+\,+}+\mathcal{L}^{(\mathrm{LC})}_{-\,-\,-\, \dots\, -\,+\,+\,+\,+}+ \dots
\Bigg\}
\,,\label{eq:Z_action1}
\end{align}
For convenience,  we shall call the new action as \emph{Z-field action} hereafter. It has the following properties: 
\begin{enumerate}[label={\it\roman*}$\,$)]
    \item There are no three point interaction vertices. The reason is that the triple-gluon vertices have been effectively resummed inside the Wilson lines. Thus, the lowest multiplicity vertex is the four-point MHV vertex.
    \item At the classical level there are no all-plus, all-minus, as well as $(-+\dots +)$, $(- \dots - +)$ vertices.
    \item There are MHV vertices, $(--+\dots +)$, corresponding to MHV amplitudes in the on-shell limit.
    \item There are $\overline{\mathrm{MHV}}$ vertices, $(-\dots - ++)$, corresponding to $\overline{\mathrm{MHV}}$ amplitudes in the on-shell limit.
    \item All vertices have the form which can be easily calculated. In the following, we discuss the general form for any vertex in the Z-field action.
\end{enumerate}

Let us now discuss a general form of the vertex. Without loosing generality, we consider the negative helicity fields adjacent and moreover focus on the color ordered vertex, defined by
\begin{equation}
    \mathcal{U}_{-\,\dots\,-\,+\,\dots\,+}^{b_{1}\dots b_{n}}\left(\mathbf{p}_{1},\dots,\mathbf{p}_{n}\right)= \!\!\sum 
 \mathrm{Tr}\left(t^{b_1}\dots t^{b_n}\right)
 \mathcal{U}\left(1^-,\dots,m^-,(m+1)^+,\dots,n^+\right)
\,,
\label{eq:Zvertex_color_decomp}
\end{equation}
where $m$ is the number of minus helicity legs. The sum is over noncyclic permutations. We shall also use the color ordered versions of the kernels in the expansions \eqref{eq:BstarZ_exp}-\eqref{eq:BbulletZ_exp}. Furthermore, we introduce a collective  index $[i,i+1,\dots,j]$ labeling the momentum, $\mathbf{p}_{i(i+1)\dots j}=\mathbf{p}_i+\mathbf{p}_{i+1}+\dots+\mathbf{p}_j$. Using this notation, the general form of the color ordered vertex can be written as \cite{kakkad2021}: 
\begin{multline}
    \mathcal{U}\left(1^-,\dots,m^-,(m\!+\!1)^+,\dots,n^+\right) = 
    \sum_{p=0}^{m-2}\sum_{q=p+1}^{m-1}\sum_{r=q+1}^{m}\\
    \mathcal{V}\left(\,[p\!+\!1,\dots,q]^-,[q\!+\!1,\dots,r]^-,[r\!+\!1,\dots,m\!+\!1]^+,(m\!+\!2)^+,\dots,(n\!-\!1)^+,[n,1,\dots,p]^+\right) \\
    \overline{ \Omega}\left(n^+,1^-,\dots,p^-\right) \,\,
    \overline{ \Psi}\left((p\!+\!1)^-,\dots,q^-\right) \,\, 
    \overline{ \Psi}\left((q\!+\!1)^-,\dots,r^-\right) \,\, 
    \overline{ \Omega}\left((r\!+\!1)^-,\dots,m^-,(m\!+\!1)^+\right) \,\,
    \label{eq:Z_gen_ker}
\end{multline}
This can be easily understood as following. The substitution of $B$ fields in terms of $Z$ fields can only multiplicate negative helicity legs. Thus we start with MHV vertex with $n-m$ positive helicity legs. Also, since we have considered all negative helicity legs adjacent, the only possible contributions are the ones where the $B$ fields is substituted to at least one of the four adjacent $(-\,-\,+\,+)$ legs in the MHV vertex. Summing over all such contribution gives (\ref{eq:Z_gen_ker}). Although the analytic formula do not seem to collapse, in general, to any simple form, the above  expression is operational and
 can be readily applied in the actual amplitude calculation. We discuss this in the following.

\subsection{Amplitudes}
\begin{figure}
    \centering
 \includegraphics[width=10cm]{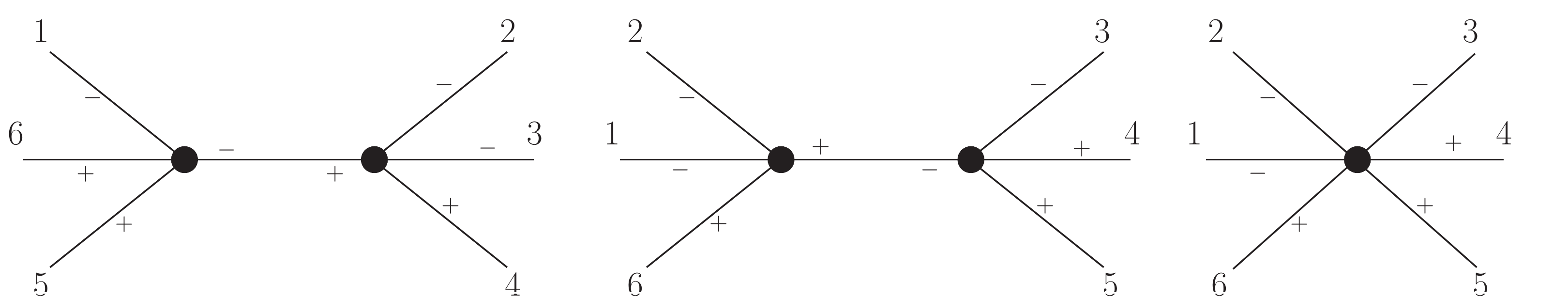}
    \caption{\scriptsize 
    Diagrams contributing to the 6-point NMHV amplitude $(---+++)$.
    }
    \label{fig:NMHV6}
\end{figure}

\begin{figure}[h]
    \centering
 \includegraphics[width=10cm]{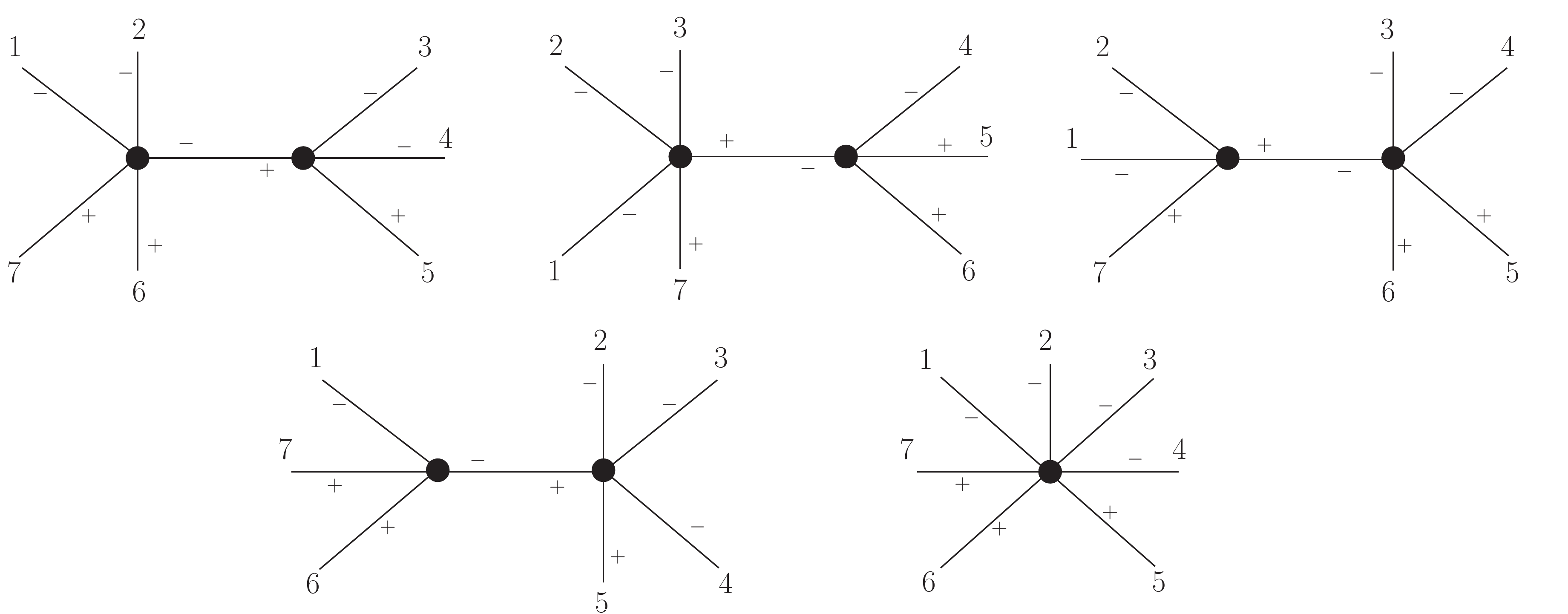}
    \caption{\scriptsize 
    Diagrams  contributing to the 7-point NNMHV amplitude $(----+++)$.
    }
    \label{fig:NNMHV7}
\end{figure}

Using the Z-field action we computed several tree-level amplitudes. The MHV and $\overline{\text{MHV}}$ vertices alone give the corresponding on-shell amplitudes. Consider the 5-point $\overline{\text{MHV}}$ vertex. It is easily obtained from \eqref{eq:Z_gen_ker}. In Fig.~\ref{fig:MHVbar5_vertex} we show the contributing terms.  In the on-shell limit, the sum of these diagrams reduces to the known formula for the $\overline{\text{MHV}}$ amplitude:
\begin{figure}
\vspace{-0.5cm}
    \centering
 \includegraphics[width=10cm]{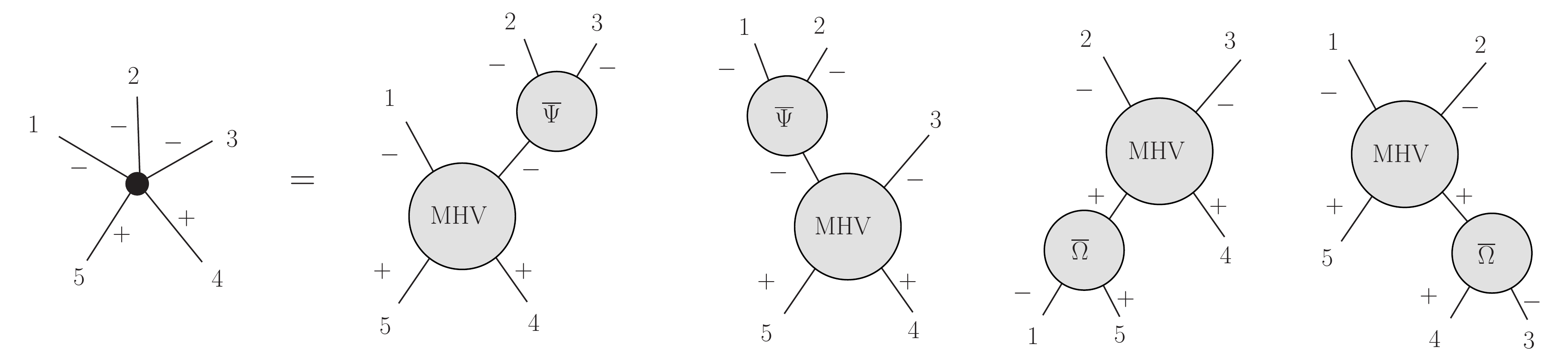}
    \caption{\scriptsize 
    The contributions to the color-ordered $\overline{\text{MHV}}$ vertex, with helicity $(---++)$.
    }
    \label{fig:MHVbar5_vertex}
\end{figure}
\begin{equation}
    \mathcal{A}(1^-,2^-,3^-,4^+,5^+) =
 g^3 \left(\frac{p_{4} ^{+}}{p_{5}^{+}}\right)^{2}
\frac{\widetilde{v}_{54}^{4}}{\widetilde{v}_{15}\widetilde{v}_{54}\widetilde{v}_{43}  \widetilde{v}_{32}\widetilde{v}_{21} } \, .
 \label{eq:5G_MHVbar_onshell}
\end{equation}
For the 6-point NMHV amplitude $(---+++)$ we have just three contributing diagrams depicted in Fig.~\ref{fig:NMHV6}. The sum of these diagrams reproduce in the on-shell limit the known result \cite{Kosower1990}.
For 7-point NNMHV amplitude $(----+++)$ we had just five contributing diagrams depicted in Fig.~\ref{fig:NNMHV7}.
Furthermore, the higher multiplicity amplitudes, up to 8-point NNMHV, were calculated and shown to be in agreement with the standard methods \cite{Dixon:2010ik}. The maximum number of diagrams we encountered 
in the latter case was 13. 

\section{Conclusions}
We developed a new action for gluodynamics  by  canonically transforming (Eq.~\eqref{eq:generatingfunc2}) the light-cone Yang-Mills action. The most striking property of the new action is that it has no triple-gluon vertex.
Consequently, the number of diagrams needed to calculate the amplitudes is greatly reduced. Also, the geometric structure of the field transformations leading to the new action is incredibly rich and requires further investigation. Finally, a formulation at loop level seems feasible \cite{Brandhuber2007a,Fu_2009, Boels_2008, Ettle2007,Brandhuber2007,Elvang_2012} and is under development.

\section*{Acknowledgements}
H.K. and P.K. are supported by the National Science Centre, Poland grant no. 2018/31/D/ST2 /02731.  A.M.S. is supported  by the U.S. Department of Energy Grant 
 DE-SC-0002145 and  in part by  National Science Centre in Poland, grant 2019/33/B/ST2/02588.

\bibliography{BiBTeX.bib}

\nolinenumbers

\end{document}